\documentclass{article}
\usepackage[margin=1in]{geometry}
\usepackage{amsmath}
\usepackage{epsfig}
\numberwithin{equation}{section} % amsmath

\newcommand{\beq}{\begin{equation}}
\newcommand{\eeq}{\end{equation}}
\newcommand{\beqa}{\begin{eqnarray}}
\newcommand{\eeqa}{\end{eqnarray}}
\newcommand{\bdm}{\begin{displaymath}}
\newcommand{\edm}{\end{displaymath}}

\newcommand{\Eq}[1]{Eq.\ (\ref{#1})}
\newcommand{\Eqs}[2]{Eqs.\ (\ref{#1}) and (\ref{#2})}
\newcommand{\Eqss}[3]{Eqs.\ (\ref{#1}), (\ref{#2}) and (\ref{#3})}

\newcommand{\Rref}[1]{Ref.\ \cite{#1}}

\newcommand{\Fig}[1]{Fig.\ \ref{#1}}
\newcommand{\Figs}[2]{Figs.\ \ref{#1} and \ref{#2}}

\newcommand{\Section}[1]{Section\ \ref{#1}}
\newcommand{\Appendix}[1]{Appendix\ \ref{#1}}
\title{
  Photon propagation in a charged Bose-Einstein condensate model
}

\author{Jos\'e F. Nieves\footnote{nieves@ltp.uprrp.edu}\\
  Laboratory of Theoretical Physics, Department of Physics\\
  University of Puerto Rico, R\'{\i}o Piedras, Puerto Rico 00936
  \and\\[12pt]
  Sarira Sahu\footnote{sarira@nucleares.unam.mx}\\
  Instituto de Ciencias Nucleares\\
  Universidad Nacional Aut\'onoma de Mexico\\
  Circuito Exterior, C. U.\\
  A. Postal 70-543, 04510 Mexico DF, Mexico\\
}

\date{November 2023}

\begin{document}
\maketitle

\begin{abstract}
  We consider the propagation of photons in a model of a
  charged scalar Bose-Einstein (BE) condensate.
  We determine the dispersion relations of the collective modes,
  as well as the photon polarization tensor and the dielectric constant
  in the model. Two modes correspond to the transverse photon polarizations,
  with dispersion relations of the usual form for
  transverse photons in a plasma. The other two modes,
  denoted as the $(\pm)$ modes, are combinations
  of the longitudinal photon and the massive scalar field. Their dispersion
  relations behave very differently as functions of momentum.
  The $(+)$ mode dispersion relation increases steadily and remains greater
  than the momentum as the momentum increases. The dispersion relation
  of the $(-)$ mode decreases in a given momentum range, with
  the group velocity being negative in that range, while
  in another range it increases steadily but remains smaller than the momentum,
  akin to the situation in a medium with an index of refraction greater than 1.
  We consider the non-relativistic limit of the $(\pm)$ dispersion
  relations and discuss some aspects of the results.
  We also determine the wavefunctions of the $(\pm)$ modes,
  which are useful to obtain the corrections to the dispersion relations,
  e.g., imaginary parts due to the damping effects and/or the effects of
  scattering, due to the interactions with the
  excitations of the system. The results can be useful in
  various physical contexts that have been considered in the literature
  involving the electrodynamics of a charged scalar BE condensate.
\end{abstract}

%end-preamble

\section{Introduction and motivation}
\label{sec:intro}

Many extensions of the standard electroweak theory depend
on the presence of additional scalar particles, both neutral and charged,
that interact with the standard model particles, in particular
with the neutrinos and photons. Such interactions typically
have implications in the context of nuclear, astrophysics,
and cosmological plasmas that may contain a background of the scalar
particles. As is well known, the intrinsic properties of neutrinos
and photons propagating in such a medium, such as the dispersion
relations and/or electromagnetic properties, are modified
in ways that lead to remarkable and observable effects.

In previous works we have carried out a systematic calculation of the neutrino
dispersion relation, including the damping and decoherence effects
in models in which the neutrinos interact with a scalar ($\phi$)
and fermion ($f$) via a coupling of the form $\bar f_R\nu_L\phi$, or just
with neutrinos themselves $\bar\nu^c_R\nu_L\phi$ (see \Rref{ns:nuphiresonance}
and references therein).
Couplings of those forms produce additional contributions to
the neutrino effective potential when the neutrino propagates
in a background of $\phi$ and $f$ particles, or a pure neutrino
background, which can effects in various neutrino physics contexts.
More recently, we have considered the problem of determining
the corresponding quantities (e.g, effective potential and/or
dispersion relation and damping) of a fermion that propagates
in a thermal background that contains a scalar Bose-Einstein (BE)
condensate\cite{ns:fbec}. The method can be applied to
the case of a neutrino propagating in such a background, and
the problem of fermions propagating in such backgrounds can be relevant
in other applications in astrophysical, heavy-ion collisions and dark-matter
contexts\cite{baym:nstar,thorsson:kaon,schmitt:kaon,li:hion,
dm:craciun,dm:sikivie,dm:fan}.

Our motivation in the present work is to consider, in a similar vein, the
propagation of photons in a thermal background that contains
a BE condensate of a charged scalar.
The problem of the propagation of photons in a medium is a classic subject
which in the language of plasma physics can be treated within the formalism
of the kinetic Boltzmann equation, and in other contexts and/or more generally
using the framework of thermal field theory (TFT).
Various aspects and applications of the problem of photon
propagation in a BE condensate of the charged scalars have been studied
in previous recent works, in various contexts and using different
approaches, for example the electrodynamics of fermions and
charged scalar bosons at low temperature\cite{dolgov},
heavy-ion collisions\cite{voskresensky}, exciton-polaritons\cite{kasprzak}, 
the BE condensation of photons\cite{kruchkov,mendonca} and the
Cerenkov effect\cite{slyu},
and the reduction of the photon velocity\cite{vestergaard},
in atomic gases.

Our goal in the present work is to extend the technique used
in \Rref{ns:fbec} to the case in which the scalar is electromagnetically
charged and study the propagation of photons in the BE condensate
background of such scalars.
The field theoretical method we used in \Rref{ns:fbec} to treat the BE
condensate, which we use here, has been discussed by various
authors\cite{weldon:phimu,filippi:phimu,schmitt:phimu,haber}.
As we will see, this method allows us to determine
succinctly the dispersion relations of the propagating modes
after the symmetry breaking associated with the BE condensation
in the model, as well as the corresponding polarization tensor and
dielectric constant. The method also
allows us to determine in a very efficient way the mixing parameters
between the longitudinal photon (plasmon) and the physical scalar field that
make up the propagating modes involving the plasmon. These are relevant
for the calculation of thermal corrections to the dispersion relations
(e.g., imaginary parts due to damping effects) and the other physical quantities
due to interactions with the thermal excitations of the BE condensate.
The results described above are obtained in a general way, within
the context of the model, but not tied to any specific application.
In this way they can be useful and pave the way for further development
in various contexts and applications such as those considered in the references
mentioned above.
While the principles and techniques used here to treat the charged
BE condensate share some common ingredients with those of \Rref{haber},
our scope, discussion and outlook for applications are very different.

The plan of the presentation describing the present work is as follows.
In \Section{sec:becmodel} we summarize the essential features
and the symmetry breaking mechanism of the scalar BE condensation model
discussed in \Rref{ns:fbec}, which serves as the basis and motivation
for the extension we present in \Section{sec:gbecmodel} to the charged
scalar case. In \Section{sec:gbecmodel} we discuss various qualitative
properties of the model that are useful for the interpretation,
such as current conservation, symmetry breaking and the spectrum.
The propagating modes and the corresponding dispersion relations are
determined in \Section{sec:dr}. As we show, two modes correspond
to the transverse photon polarizations, and their dispersion relations
have the usual form of the transverse photons in a plasma. The other
two modes, which we denote as the $(\pm)$ modes, are combinations of
the longitudinal photon and the massive scalar field.
Various distinguishing aspects of the results obtained are discussed in
\Section{sec:discussion}. While the dispersion relation of the $(+)$ mode
increases steadily and remains greater than the momentum as the
momentum increases, the dispersion relation of the $(-)$ mode
reveals some special features. A particularly noticeable one is the fact that
it decreases in a given momentum range, and the
corresponding group velocity is negative in that range. In another range, it
increases steadily for values of the momentum above a given threshold,
but such that it remains smaller than the momentum,
reminiscent of a medium with index of refraction greater than 1.
We also consider the non-relativistic limit of the dispersion
relations and discuss some aspects of the results obtained.
In particular, under some conditions, the dispersion
relation of the $(-)$ mode reduces to the expression
for the dispersion relation that has been obtained in
non-relativistic models of the BE condensation of charged
scalars\cite{2prb,3prb}.
In \Section{sec:poltensor} we obtain the expression
for the photon polarization tensor, or self-energy. On one hand,
this provides an alternative way to obtain the dispersion relations.
More importantly, it has the virtue that it gives
the contribution to the dielectric function of the system due to the
BE condensate, which is useful on its own merits, for example for studying
also the response of the system to external fields. Finally
in \Section{sec:mixing} we determine the mixing parameters of the 
spinors of the $(\pm)$ modes. These parameters can be used to obtain
the corrections to the
dispersion relations (e.g., imaginary parts due the damping effects)
and/or the effects of scattering, due to the interactions with
the excitations of the system.
In \Section{sec:conclusions} we summarize our conclusions and outlook.

\section{Scalar BE condensate model}
\label{sec:becmodel}

As a preamble to the presentation of the model for the charged
scalar coupled to the photon, we first summarize the essential features
of the scalar BE condensation model as well as its justification.

In \Rref{ns:fbec} we used the following model for the $\phi$,
\beq
\label{Lphi}
L_\phi = (d^\mu\phi)^\ast(d_\mu\phi) - V_{\phi}\,,
\eeq
where
\beq
\label{D}
d_\mu \equiv \partial_\mu - i\mu u_\mu\,,
\eeq
and
\beq
\label{Vphi}
V_\phi = m^2\phi^\ast\phi + \lambda(\phi^\ast\phi)^2\,.
\eeq
$\mu$ is a parameter, to be identified with the chemical potential
of the $\phi$ and $u^\mu$, which is to be identified with
the velocity four-vector of the medium, is
\beq
\label{u}
u^\mu = (1,\vec 0)\,,
\eeq
in the medium's own rest frame. To simplify the notation, we write
\beq
\label{vu}
v_\mu = \mu u_\mu\,.
\eeq

Expanding the $d$ term in \Eq{Lphi}, the Lagrangian density is
\beq
\label{Lmu}
L_\phi = (\partial^\mu\phi)^\ast(\partial_\mu\phi) +
i[\phi^\ast (v\cdot\partial\phi) - (v\cdot\partial\phi^\ast)\phi] - U(\phi)\,,
\eeq
where
\beq
\label{U}
U \equiv V_\phi - \mu^2\phi^\ast\phi =
-(\mu^2 - m^2)\phi^\ast\phi + \lambda(\phi^\ast\phi)^2\,.
\eeq
If $m^2 > \mu^2$, this $U$ corresponds
to a standard massive complex scalar with mass $m^2 - \mu^2$.
On the other hand, if $\mu^2 > m^2$, the minimum of the potential is
not at $\phi = 0$, and therefore $\phi$ develops a non-zero expectation
value and the $U(1)$ symmetry is broken.

We assume the second option,
\beq
\label{sbcondition}
\mu^2 > m^2\,,
\eeq
and proceed accordingly. Namely, we put
\beq
\label{phi}
\phi = \frac{1}{\sqrt{2}}\left(\phi_0 + \phi_1 + i\phi_2\right)\,,
\eeq
where
\beq
\langle\phi\rangle \equiv \frac{1}{\sqrt{2}}\phi_0\,,
\eeq
is chosen to be the minimum of
\beq
U_0 = -\frac{1}{2}(\mu^2 - m^2)\phi^2_0 + \frac{1}{4}\lambda\phi^4_0\,.
\eeq
Thus,
\beq
\label{phi0}
\phi^2_0 = \frac{\mu^2 - m^2}{\lambda}\,.
\eeq
Substituting \Eqs{phi}{phi0} in \Eq{Lphi} we obtain the
Lagrangian density for the scalar excitations $\phi_{1,2}$,
\beq
L_\phi = \frac{1}{2}
\left[(\partial^\mu\phi_1)^2 + (\partial^\mu\phi_2)^2\right] +
\phi_2 v\cdot\partial\phi_1 - \phi_1 v\cdot\partial\phi_2 - U(\phi)\,,
\eeq
where
\beq
U(\phi) = -\frac{1}{2}(\mu^2 - m^2)[(\phi_0 + \phi_1)^2 + \phi^2_2]
+ \frac{1}{4}\lambda[(\phi_0 + \phi_1)^2 + \phi^2_2]^2\,.
\eeq
Using \Eq{phi0}, the quadratic part of $U$ involves only $\phi_1$,
\beq
U_2 = \frac{1}{2} m^2_1\phi^2_1\,,
\eeq
where
\beq
m^2_1 = 2(\mu^2 - m^2)\,,
\eeq
while $\phi_2$ does not appear. However,
$\phi_1$ and $\phi_2$ are mixed by the $v^\mu$ term.
Therefore the next step is to find the propagator matrix of the $\phi_{1,2}$
complex and find the modes that have a definite dispersion relation.

A convenient way to do it is to use matrix notation,
\beq
\hat\phi = \left(
\begin{array}{l}
  \phi_1 \\ \phi_2
\end{array}\right)\,.
\eeq  
The Lagrangian density, in momentum space, can then be written in the form
\beq
L^{(2)}_\phi(k) = \frac{1}{2}
\hat\phi^\ast(k)\Delta^{-1}_\phi(k)\hat\phi(k)\,,
\eeq
where
\beq
\label{Deltainv}
\Delta^{-1}_\phi(k) = \left(
\begin{array}{ll}
  k^2 - m^2_1 & 2iv\cdot k\\
  -2iv\cdot k & k^2
\end{array}
\right)\,.
\eeq
The classical equations of motion are then
\beq
\label{becmodeleqmotion}
\Delta^{-1}_\phi(k)\hat\phi = 0\,.
\eeq
The dispersion relations of the eigenmodes are given by the solutions of
\beq
D = 0\,,
\eeq
where $D$ is the determinant of $\Delta^{-1}_\phi$,
\beq
\label{D2}
D = (\omega^2 - \kappa^2)(\omega^2 - \kappa^2 - m^2_1) - 4\mu^2\omega^2\,.
\eeq
In \Eq{D2} we have used \Eq{vu} and introduced the variables
$\omega$ and $\kappa$,
\beqa
\label{omegakappadef}
\omega & = & u\cdot k\,,\nonumber\\
\kappa & = & \sqrt{\omega^2 - k^2}\,.
\eeqa
In the rest frame of the medium
\beq
\label{krestframe}
k^\mu = (\omega,\vec\kappa)\,,
\eeq
with $\kappa = |\vec\kappa|$. The dispersion relations obtained in this way
can be written in the form
\beq
\label{omegapmfinal}
\omega^2_{\pm}(\kappa) = \kappa^2 + \frac{1}{2} m^2_{\rho} \pm
\sqrt{\frac{1}{4} m^4_{\rho} + 4\mu^2\kappa^2}\,,
\eeq
where
\beq
\label{becmodelmasses}
m^2_{\rho} = m^2_1 + 4\mu^2\,,
\eeq
and they satisfy
\beqa
\label{omegapmzerokappa}
\omega_{+}(0)  & = & m_\rho\,,\nonumber\\
\omega_{-}(0)  & = & 0\,.
\eeqa
The combinations of $\phi_{1,2}$ that have the definite dispersion
relations $\omega_{\pm}$ can be obtained from \Eq{becmodeleqmotion}.
The mode corresponding to $\omega_{-}$
is the realization of the Goldstone mode associated
with the breaking of the global $U(1)$ symmetry.
However, its dispersion relation is not the usual $\omega = \kappa$,
but the one given by $\omega_{-}(\kappa)$ in \Eq{omegapmfinal}.

As discussed in \Rref{ns:fbec},
the motivation for considering this model is based on the following result:
the calculation of the effective potential $V_\text{eff}(T,\mu)$ for $\phi$
can be carried out in TFT using $\mu = 0$ in the partition
(and/or distribution) function, but using the
$\mu$-dependent Lagrangian density $L_\phi$ defined in \Eq{Lphi}.
Then neglecting the $T$-dependent terms (that is, at zero temperature),
$V_\text{eff}(0,\mu)$ is simply the $U$ potential given in \Eq{U}.
In other words, calculations with the $\phi$ propagators, 
$\phi$ Lagrangian density and partition function for $\phi$,
\begin{equation}
Z = e^{-\beta H + \alpha Q}\,,
\end{equation}
give the same result as calculating them with the Lagrangian density
for $\phi^\prime$ obtained by using
$\partial_\mu \rightarrow D_\mu = \partial_\mu - i\mu u_\mu$ and using
\beq
Z = e^{-\beta H^\prime}\,.
\eeq

The simple way to obtain this result is the following.
The evolution equation for $\phi$ is
\beq
\label{hamiltoneqphi}
i\partial_t\phi = - [\cal H,\phi]\,.
\eeq
If now calculate the time derivative of $\phi^\prime \equiv e^{i\mu t}\phi$
we get
\beqa
\partial_t\phi^\prime & = & i\mu\phi^\prime + e^{i\mu t}\partial_t\phi
\nonumber\\
& = & i\mu\phi^\prime + e^{i\mu t}i[\cal H,\phi]\nonumber\\
& = & i\mu\phi^\prime + i[\cal H,\phi^\prime]\nonumber\\
& = & -i[\mu Q,\phi^\prime] + i[\cal H,\phi^\prime]\nonumber\\
& = & i[\cal H - \mu Q,\phi^\prime]\,.
\eeqa
Therefore
\beq
i\partial_t\phi^\prime = -[\cal H - \mu Q,\phi^\prime]\,,
\eeq
which is the result stated above, with the identification of $\cal H^\prime$.
On the other hand, when we express the partition function in terms
of the field $\phi^\prime$, then
\beq
Z \equiv e^{-\beta\cal H + \alpha Q} = e^{-\beta(\cal H - \mu Q)} =
e^{-\beta\cal H^\prime}\,.
\eeq
That is, in the calculations involving $\phi^\prime$
we use the partition function with $\cal H^\prime$ and zero chemical
potential.

The operator proof we have presented
seems to be applicable to any field (e.g., a fermion field),
not involving Lagrangian dynamics arguments, and therefore
the result shown above holds for any field and not just for
a scalar field. On the basis of this result, the Lagrangian formulation
that we have presented is justified, which is
the most efficient way to do the dynamics
of the $\phi^\prime$, that is finding the evolution equation to obtain the
dispersion relations and other properties (e.g., the propagators)
of the eigenmodes.

\section{Charged scalar BE condensate model}
\label{sec:gbecmodel}

\subsection{Lagrangian}

We extend the model by assuming that the $\phi$ field is electrically
charged and interacts with electromagnetic field in the usual way.
The Lagrangian density is then
\beq
L = -\frac{1}{4}F^2 + L_{\phi A}\,,
\eeq
where $F_{\mu\nu}$ is the electromagnetic field tensor and
\beq
\label{Lphigauged}
L_{\phi A} = (\hat D^\mu\phi)^\ast(\hat D_\mu\phi) -  V_\phi\,,
\eeq
with\cite{footnote1}
\beq
\label{Dgauged}
\hat D_\mu = d_\mu + iqA_\mu = \partial_\mu - iv_\mu + iqA_\mu\,,
\eeq
and $V_\phi$ given as before in \Eq{Vphi}. $v^\mu$ is defined in \Eq{vu} and
we will sometimes write
\beq
\hat D_\mu = \partial_\mu -i \omega_\mu\,,
\eeq
where
\beq
\omega_\mu = v_\mu - qA_\mu\,.
\eeq

\subsection{Spectrum of the model}

To determine the spectrum of the model we adopt the unitary gauge,
which is the convenient one to use for this purpose.
Thus we parametrize $\phi$ in the form
\beq
\phi = \frac{1}{\sqrt{2}}(\phi_0 + \rho)e^{i\theta/\phi_0}\,.
\eeq
The field $\theta$ does not appear in $V_\phi$, and
by a gauge transformation it disappears also from the kinetic term.
To be clear that we are employing the unitary gauge we will denote
by $V_\mu$ the transformed vector potential
$V_\mu = A_\mu + \frac{1}{q\phi_0}\partial_\mu\theta$.
It then follows that
\beq
\label{LAphi4}
L_{\phi A}  = \frac{1}{2}(\partial\rho)^2 +
\frac{1}{2}(v - qV)^\mu(v - qV)_\mu(\phi_0 + \rho)^2 - V_\phi\,,
\eeq
or, expanding the various factors,
\beq
\label{LAphi3}
L_{\phi A} = \frac{1}{2}(\partial\rho)^2 + \frac{1}{2}(q\phi_0)^2 V^2 +
\frac{1}{2}q^2 V^2\rho^2 + q^2\phi_0\rho V^2 -
q v\cdot V(\phi_0 + \rho)^2 - U_\phi\,,
\eeq
with
\beq
\label{Uphigbec}
U_\phi = -\frac{1}{2}(\mu^2 - m^2)(\phi_0 + \rho)^2 +
\frac{1}{4}(\phi_0 + \rho)^4\,.
\eeq

Proceeding as in \Section{sec:becmodel}, we assume
\beq
\mu^2 > m^2\,,
\eeq
and choose $\phi_0$ such that $U_\phi$ has the minimum at $\rho = 0$, thus,
\beq
\phi^2_0 = \frac{\mu^2 - m^2}{\lambda}\,.
\eeq
From \Eq{LAphi3}, the bilinear part of $L$, including the $F^2$ term, is then
\beq
\label{Lbilinearx}
L^{(2)} = -\frac{1}{4}F^2 + \frac{1}{2}m^2_V V^2 +
\frac{1}{2}(\partial\rho)^2 - \frac{1}{2}m^2_1\rho^2 -
2m_V (v\cdot V)\rho\,,
\eeq
where
\beqa
\label{massparameters}
m_V & = & q\phi_0\,,\nonumber\\
m^2_1 & = & 2(\mu^2 - m^2)\,.
\eeqa
The physical picture that emerges is this:
the field $\theta$ becomes the longitudinal component of $V$, and
we end up with two fields, $V$ and $\rho$. The longitudinal component of
$V$ is mixed with the scalar field $\rho$ (the term $\rho v\cdot V)$.
As we show, the consequence is that the propagating modes with definite
dispersion relations involve a superposition of the longitudinal component
of $V$ and the $\rho$.

The following comment is in order before proceeding.
In \Eq{LAphi3} there is a term linear in $V_\mu$ of the form  
$q\phi^2_0 v\cdot V = (q\mu\phi^2_0)u\cdot V$.
This term is a tadpole, meaning that the ground state has a net
contribution to the charge due to $\phi_0$. Indeed, the
definition of the electromagnetic current,
\beq
j^{(em)}_\mu = -\frac{\partial L_{\phi A}}{\partial A^\mu}\,,
\eeq
gives
\beq
\label{jem}
j^{(em)}_\mu = iqJ_\mu\,,
\eeq
where
\beq
\label{J}
J_\mu = \phi^\ast\hat D_\mu\phi - (\hat D_\mu\phi)^\ast\phi\,,
\eeq
and it is straightforward to verify that $j^{(em)}_\mu$ is conserved,
as it should be. It then follows that
the symmetry breaking mechanism produces a net current density given by
\beq
\label{jemnet}
j^{(em)}_\mu = q\mu\phi^2_0 u_\mu\,.
\eeq
Physically, the total charge of the system must be zero, which
is assured by assuming that the $\phi$ background is superimposed
on an oppositely charged background to render the total system electrically
neutral. Thus for practical purposes we can disregard that tadpole term.

\section{Eigenmodes and the dispersion relations}
\label{sec:dr}

\subsection{The evolution equations}

We write it in momentum space by making the correspondence\cite{footnote2},
\beqa
\label{xkcorrespondence}
-\frac{1}{4}F^2 & \rightarrow & -V^{\ast\mu}k^2 \tilde g_{\mu\nu} V^\nu\,,
\nonumber\\
\frac{1}{2}(\partial \rho)^2 & \rightarrow & k^2\rho^\ast\rho\,,\nonumber\\
\frac{1}{2}m^2_V V^2 & \rightarrow & m^2_V V^{\ast\mu}V_\mu\,,\nonumber\\
\frac{1}{2}m^2_1 \rho^2 & \rightarrow & m^2_1 \rho^{\ast}\rho\,,
\nonumber\\
(u\cdot V)\rho & \rightarrow & (u\cdot V)^\ast\rho + c.c\,,
\eeqa
and therefore
\beq
\label{Lbilineark}
L^{(2)}(k) = -V^{\ast\mu}[k^2\tilde g_{\mu\nu} - m^2_V g_{\mu\nu}]V^\nu +
\rho^\ast(k^2 - m^2_1)\rho -
2\mu m_V[(u\cdot V)^\ast\rho + c.c]\,,
\eeq
where
\beq
\tilde g_{\mu\nu} = g_{\mu\nu} - \frac{k_\mu k_\nu}{k^2}\,.
\eeq
If it were not for the last term, this is the usual free Lagrangian density
for a vector with squared mass $m^2_V$ and a scalar with squared mass $m^2_1$.
However the last term mixes the scalar with the longitudinal
components of $V^\mu$.

We decompose $V^\mu$ in the form
\beq
\label{VTLk}
V^\mu = V^\mu_T + V_L e^\mu_{3} + \frac{k^\mu}{\sqrt{k^2}}V_k\,,\qquad
V^\mu_T = \sum_{i= 1,2}V_i e^\mu_i
\eeq
where
\beqa
e^\mu_{1,2} & = & (0,\vec e_{1,2})\,,\nonumber\\
e^\mu_3 & = & \frac{\tilde u^\mu}{\sqrt{-\tilde u^2}}\,,
\eeqa
with
\beqa
\label{utilde2}
\tilde u_\mu =
\tilde g_{\mu\nu} u^\nu & = & u_\mu - \frac{(k\cdot u)k_\mu}{k^2}\,,\nonumber\\
\sqrt{-\tilde u^2} & = & \frac{\kappa}{\sqrt{k^2}}\,.
\eeqa
In particular,
\beqa
\tilde u\cdot V & = & -\sqrt{-\tilde u^2} V_L\,,\nonumber\\
k\cdot V & = & \sqrt{k^2}V_k\,,
\eeqa
and therefore
\beq
u\cdot V = \tilde u\cdot V + \frac{(k\cdot u)(k\cdot V)}{k^2}
= -\sqrt{-\tilde u^2} V_L + \frac{(k\cdot u)V_k}{\sqrt{k^2}}\,.
\eeq

Thus the quadratic part of the Lagrangian density is
\beqa
\label{L2withVk}
L^{(2)}(k) & = & -(k^2 - m^2_V)V^\ast_T\cdot V_T +
(k^2 - m^2_V)V^{\ast}_L V_L + m^2_V V^\ast_k V_k +
(k^2 - m^2_1)\rho^\ast\rho\nonumber\\
&&\mbox{} + 2\mu m_V \sqrt{-\tilde u^2}\left[V^\ast_L\rho + c.c.\right] -
\frac{2\mu m_V (k\cdot u)}{\sqrt{k^2}}\left[V^\ast_k\rho + c.c\right]\,.
\eeqa
Since the component $V_k$ does not have a kinetic energy term, and therefore
it is not a dynamical variable, one way to proceed is to
eliminate it from the Lagrangian by using its Lagrange equation.
The equation is
\beq
m^2_V V_k - \frac{2\mu m_V(k\cdot u)}{\sqrt{k^2}}\rho = 0\,,
\eeq
and therefore
\beq
\label{Vk}
V_k = \frac{2\mu(k\cdot u)}{m_V\sqrt{k^2}}\rho\,.
\eeq
Substituting this back in $L^{(2)}(k)$ we then get
\beqa
\label{L2withoutVk}
L^{(2)}(k) & = & -(k^2 - m^2_V)V^\ast_T\cdot V_T +
(k^2 - m^2_V)V^{\ast}_L V_L +
\left(k^2 - m^2_1 - \frac{4\mu^2(k\cdot u)^2}{k^2}\right)\rho^\ast\rho
\nonumber\\
&&\mbox{} + 2\mu m_V \sqrt{-\tilde u^2}\left[V^\ast_L\rho + c.c.\right]\,.
\eeqa
The alternative way is to find the equations for $V_L$ and $\rho$
and eliminate $V_k$ using \Eq{Vk}. The final equations for $V_L$ and
$\rho$ should be the same. As a verification,
the Lagrange equations from \Eq{L2withVk} are,
\beqa
(k^2 - m^2_V)V_L  + 2\mu m_V\sqrt{-\tilde u^2}\rho & = & 0\,,\nonumber\\
(k^2 - m^2_1)\rho + 2\mu m_V\sqrt{-\tilde u^2}V_L -
2\mu m_V\frac{(k\cdot u)}{\sqrt{k^2}}V_L & = & 0\,,
\eeqa
and using \Eq{Vk} they become
\beqa
\label{VLrhoeqs}
(k^2 - m^2_V)V_L  + 2\mu m_V\sqrt{-\tilde u^2}\rho & = & 0\,,\nonumber\\
2\mu m_V\sqrt{-\tilde u^2}V_L +
\left[k^2 - m^2_1 - \frac{4\mu^2(k\cdot u)^2}{k^2}\right]\rho & = & 0\,,
\eeqa
which is the same that is obtained from \Eq{L2withoutVk}.
The picture that emerges is that the transverse
modes develop a standard degenerate dispersion relation for the two
polarizations, while the longitudinal mode $V_L$ mixes with the scalar $\rho$.
Our next task is to solve the evolution equations we have obtained,
to find the dispersion relations and the corresponding
mixing parameters $V_L$ and $\rho$ of the eigenmodes.

\subsection{Dispersion relations}

Using \Eq{utilde2}, it follows that at $\kappa = 0$
the equations in \Eq{VLrhoeqs} decouple,
\beqa
(\omega^2 - m^2_V)V_L & = & 0\,,\nonumber\\
(\omega^2 - m^2_{\rho})\rho & = & 0\,,
\eeqa
where
\beq
\label{rhoeffectivemass}
m^2_{\rho} = m^2_1 + 4\mu^2\,.
\eeq
Thus we can identify the plasma frequency as
\beq
\label{Veffectivemass}
\omega^2_{p0} = m^2_V = q^2 \phi^2_0 = \frac{q^2(\mu^2 - m^2)}{\lambda}\,,
\eeq
while the $\rho$ mode has mass $m^2_{\rho}$,
which coincides with the mass of the ``$+$'' mode in the un-gauged model
(see \Eq{becmodelmasses}).

For $\kappa \not= 0$ the dispersion relations are the solutions of
\beq
\label{dreq1}
(k^2 - m^2_V)\left[k^2 - m^2_1 - \frac{4\mu^2(k\cdot u)^2}{k^2}\right] -
\frac{4\mu^2 m^2_V\kappa^2}{k^2} = 0\,,
\eeq
which using \Eq{omegakappadef} can be written as
\beq
(k^2 - m^2_{\rho})(k^2 - m^2_V) - 4\mu^2\kappa^2 = 0\,.
\eeq
This is a quadratic equation for $k^2$, with solutions
\beq
\label{gbecdr}
\omega^2 = \kappa^2 +
\frac{1}{2}(m^2_{\rho} + m^2_{V}) \pm \left[
\frac{1}{4}(m^2_{\rho} - m^2_{V})^2 + 4\mu^2\kappa^2\right]^{\frac{1}{2}}\,.
\eeq

\section{Discussion}
\label{sec:discussion}

\begin{figure}
\begin{center}
\epsfig{file=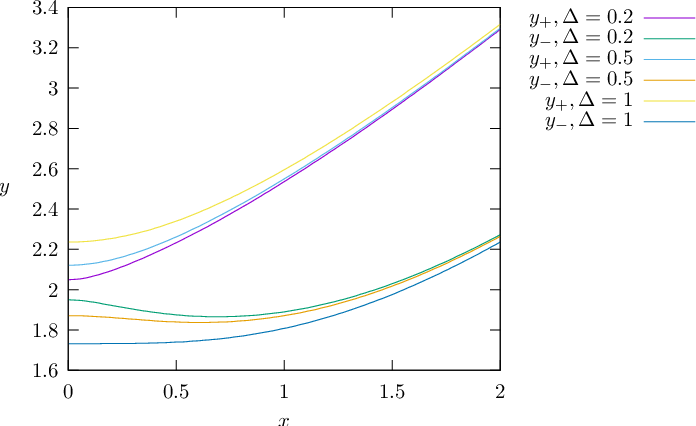,bbllx=150,bblly=294,bburx=484,bbury=499}
\end{center}
\caption[] {
  Sample plots of the dispersion relations as written in \Eq{gbecdrnorm}.
  We have set $M = 2$, and $\Delta$ as indicated. The upper branches
  correspond to $\omega_{+}$ and the lower ones to $\omega_{-}$.
  The lowest branch, corresponding to $\omega_{-}$ with $\Delta = 1$,
  is well approximated by a formula of the form given in \Eq{drprb}
  for small values of $\kappa$. See the discussion in the text.
  \label{fig:dr}
}
\end{figure}

For the following discussion it is convenient to normalize the
various parameters as follows,
\beqa
x^2 & = & \frac{\kappa^2}{2\mu^2}\,,\nonumber\\
y^2 & = & \frac{\omega^2}{2\mu^2}\,,\nonumber\\
M^2 & = & \frac{m^2_{\rho} + m^2_V}{4\mu^2}\,,\nonumber\\
\Delta & = & \frac{|m^2_{\rho} - m^2_V|}{4\mu^2}\,.
\eeqa
In terms of them, the dispersion relations given in \Eq{gbecdr}
are expressed in the form
\beq
\label{gbecdrnorm}
y^2_{\pm} = x^2 + M^2 \pm \sqrt{\Delta^2 + 2x^2}\,.
\eeq
For guidance, plots of the dispersion relations are shown
in \Fig{fig:dr} for sample values of the mass parameters.
It must be kept in mind that numerically
$\Delta < M^2$, and also it is important to recognize that
\beq
\label{M2gt1}
M^2 > 1\,.
\eeq
\Eq{M2gt1} follows from the formulas for $m^2_{V}$ and $m^2_{\rho}$
in terms of $\mu$ (e.g. \Eqs{Veffectivemass}{rhoeffectivemass}),
that give
\beq
M^2 = 1 + \frac{1}{4\mu^2}\left[
  \frac{q^2(\mu^2 - m^2)}{\lambda} + 2(\mu^2 - m^2)\right]\,.
\eeq
Remembering \Eq{sbcondition}, this implies \Eq{M2gt1}.

\begin{figure}
\begin{center}
\epsfig{file=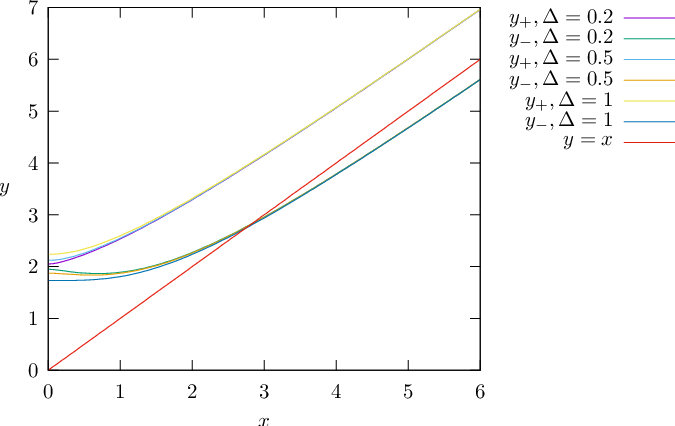,bbllx=160,bblly=294,bburx=484,bbury=499}
\end{center}
\caption[] {
  Sample plots of the dispersion relations to illustrate
  that $\omega_{-} < \kappa$ when $\kappa$ is sufficiently large;
  i.e., \Eqs{highkappax}{highkappa}.
  For reference the line $y = x$ is shown as well.
  We have set $M = 2$, and $\Delta$ as indicated.
  \label{fig:drmhigh}
}
\end{figure}

From \Eq{gbecdrnorm} it follows that $y_{+}$ increases steadily
as $x$ increases, and $y_{+} > x$, for all values of $x$.
In other words $\omega_{+}$ increases steadily
with $\kappa$, and $\omega_{+} > \kappa$, for all values
of $\kappa$. However, $\omega_{-}$ has some distinguishing features,
and its behavior is different depending on the range of the value of
$\kappa$. The plots in \Figs{fig:drmhigh}{fig:drmlow} show
the dispersion relations covering different regimes of $\kappa$
to emphasize some of these features.

Firstly, it follows from \Eq{gbecdrnorm} that
\beq
y_{-} < x
\eeq
for
\beq
\label{highkappax}
x^2 > \frac{1}{2}(M^4 - \Delta^2)\,.
\eeq
That is,
\beq
\omega_{-} < \kappa\,,
\eeq
for
\beq
\label{highkappa}
\frac{\kappa^2}{\mu^2} > M^4 - \Delta^2\,,
\eeq
as illustrated in \Fig{fig:drmhigh}. The dispersion relation in the
$\kappa$ regime indicated in \Eq{highkappa}
is reminiscent of the photon dispersion relation in a medium
with an index of refraction greater than 1.

\begin{figure}
\begin{center}
\epsfig{file=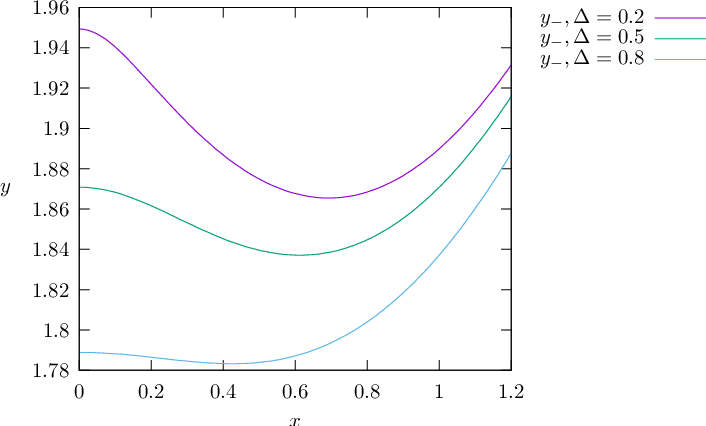,bbllx=145,bblly=294,bburx=484,bbury=499}
\end{center}
\caption[] {
  Sample plots of the dispersion relation for $\omega_{-}$
  illustrating the cases in which it has a miminum
  for certain range of values of the mass parameters.
  We have set $M = 2$, and $\Delta$ as indicated.
  \label{fig:drmlow}
}
\end{figure}
Secondly, another noteworthy feature is the fact that if $\Delta < 1$,
then $\omega_{-}$ decreases in a certain range of $\kappa$,
and reaching a minimum value.
This is illustrated in \Fig{fig:dr} in the plots of $\omega_{-}$ for
$\Delta = 0.2, 0.5$, and more emphatically in \Fig{fig:drmlow}.
In any case, the expression for $\omega^2_{-}$, as given by
\Eq{gbecdrnorm}, remains positive for all values of $\kappa$. To show this
let us note the following.
From \Eq{gbecdrnorm} it follows that $y^2_{-}$ becomes zero
at $x = x_{\pm}$, where
\beq
\label{kappapm}
x^2_{\pm} = (1 - M^2) \pm
\sqrt{(1 - M^2)^2 - (M^4 - \Delta^2)}\,.
\eeq
Since $M^2 > 1$, as shown in \Eq{M2gt1}, there is no solution of the equation
\beq
y^2_{-} = 0\,,
\eeq
and $\omega^2_{-}$ remains positive for all values of $\kappa$.

A related feature, in the case that $\Delta < 1$, is that
the group velocity of the $\omega_{-}$ branch becomes negative
in the range of $\kappa$ in which $\omega_{-}$ decreases.
\begin{figure}
\begin{center}
\epsfig{file=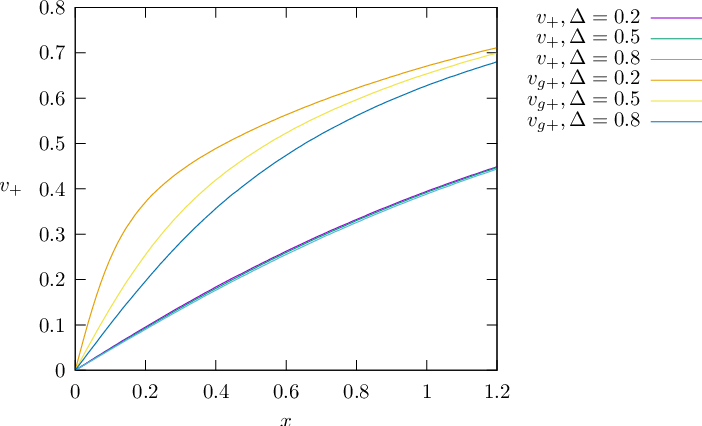,bbllx=147,bblly=294,bburx=484,bbury=499}
\end{center}
\caption[] {
  Sample plots of the group velocity $v_{g+}$ for the
  $\omega_{+}$ branch as written in \Eq{vg}.
  We have set $M = 2$, and $\Delta$ as indicated.
  \label{fig:vp}
}
\end{figure}
\begin{figure}
\begin{center}
\epsfig{file=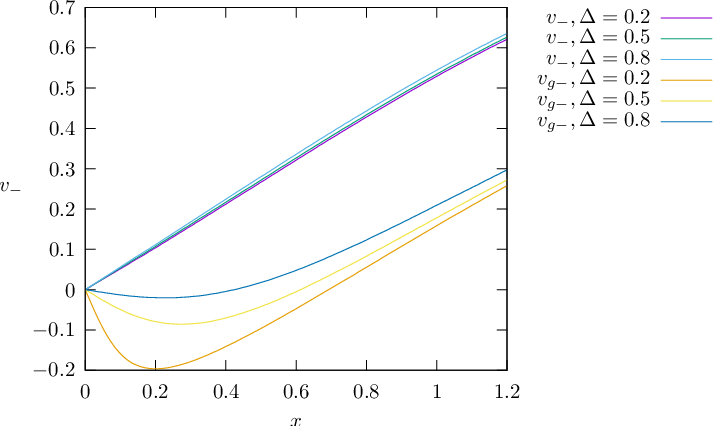,bbllx=142,bblly=294,bburx=484,bbury=499}
\end{center}
\caption[] {
  Sample plots of the group velocity $v_{g-}$ for the
  $\omega_{-}$ branch as written in \Eq{vg}.
  We have set $M = 2$, and $\Delta$ as indicated.
  \label{fig:vm}
}
\end{figure}
The group velocity for each branch is
\beq
\label{vg}
v_g = \frac{\partial\omega}{\partial\kappa} =
\frac{x}{y}\left[
1 \pm \frac{1}{\sqrt{\Delta^2 + 2x^2}}\right]\,.
\eeq
Plots of $v_{g}$ are shown in \Figs{fig:vp}{fig:vm} for the
$\omega_{\pm}$ branches, respectively. Each figure
also displays the corresponding plot of $v_{\pm} \equiv \kappa/\omega_{\pm}$
for reference. In each branch, $v$ does not change appreciably
for the different values of $\Delta$.
The group velocity of the $\omega_{-}$ branch becomes negative in
the range of $\kappa$ in which $\omega_{-}$ decreases. In fact
it is zero at
\beq
x = \sqrt{\frac{1 - \Delta^2}{2}}\,,
\eeq
and it is negative (positive) for $\kappa$ smaller (larger) than
this value, respectively.

One particular feature worth mentioning is the result for the non-relativistic
(NR) limit of the $\omega_{-}$ branch.
For sufficiently small values of $\kappa$, the dispersion relations are
\beq
\label{gbecdrnormnr}
(y^2_{\pm})_{NR} = (M^2 \pm \Delta) + \left(1 \pm \Delta^{-1}\right)x^2
\mp \frac{x^4}{2\Delta^3}\,.
\eeq
Thus, in particular,
\beq
\label{gbecdrnormmNR}
(y^2_{-})_{NR} = (M^2 - \Delta) + \left(1 - \Delta^{-1}\right)x^2
+ \frac{x^4}{2\Delta^3}\,.
\eeq
Moreover, in the case that $\Delta \simeq 1$, the dispersion of
the $(-)$ branch is
\beq
\label{drprb}
(y^2_{-})_{NR} \simeq (M^2 - 1) + \frac{1}{2}x^4\,.
\eeq
To state this more precisely, recall that in the NR limit,
the chemical potential, which we denote by $\mu_{nr}$, is related
to the chemical potential that we use in TFT by
\beq
\mu = m + \mu_{nr}\,.
\eeq
This follows from the usual form of the distribution functions,
\beq
f(p) = \frac{1}{e^{\beta(E(p) - \mu)} \pm 1}\,,
\eeq
and from the NR counterpart
\beq
f_{NR}(p) = \frac{1}{e^{\beta(p^2/2m - \mu_{nr})} \pm 1}\,.
\eeq
Let us consider specifically the case
\beq
m \gg \kappa, \mu_{nr}, m_V\,.
\eeq
In that limit \Eq{gbecdrnormmNR} gives
\beq
(y^2_{-})_{NR} = \frac{m^2_V}{2\mu^2} +
x^2 O\left(\frac{\mu_{nr}}{m},\frac{m^2_V}{m^2}\right) + 
\frac{1}{2}x^4\left(1 +
O\left(\frac{\mu_{nr}}{m},\frac{m^2_V}{m^2}\right)\right)\,,
\eeq
or, in terms of the physical variables,
\beq
(\omega^2_{-})_{NR} = m^2_V + \left(\frac{\kappa^2}{2m}\right)^2 +
O\left(\frac{\kappa^2\mu_{nr}}{m},\frac{\kappa^2 m^2_V}{m^2}\right)\,.
\eeq
Thus if the $O\left(\kappa^2\mu_{nr}/m,\kappa^2 m^2_V/m^2\right)$ terms can be
neglected, the $\kappa^2$ term does not appear, which is reminiscent of
the dispersion relation that has been obtained in NR models of the
BE condensation of charged scalars\cite{2prb}. However, in our case,
the $(\pm)$ dispersion relations are different, and this occurs only
for the $\omega_{-}$ dispersion relation. More generally,
the presence of a $O(\kappa^2)$ term (even in the $T = 0$
regime that we are considering) is mentioned in \Rref{3prb}, as well as the
fact that the coefficient is negative; which in our
case can occur for $\omega_{-}$ if $\Delta < 1$.

\section{Polarization tensor and dielectric constant}
\label{sec:poltensor}

An alternative way to obtain the dispersion relations is through
the photon polarization tensor, or self-energy.  This method has also the
virtue that it provides us with an expression for the dielectric
function of the system, which is useful on its own merits.
The starting point is the bilinear part of the Lagrangian density for
$V_\mu$ and $\rho$ given in \Eq{L2withoutVk}.
The idea is to \emph{integrate out} the variable $\rho$,
which is the tree-level analog of calculating the 1-loop self-energy,
leaving only the dynamical variables of $V$, and identify the
polarization tensor by comparing the result with
\beq
L_{eff} = -V^{\ast\mu}\left[(k^2 - \pi_T)R_{\mu\nu} +
(k^2 - \pi_L)Q_{\mu\nu}\right]V^\nu\,,
\eeq
where $R_{\mu\nu}$ and $Q_{\mu\nu}$ are defined by
\beqa
Q_{\mu\nu} & = & \frac{\tilde u_\mu \tilde u_\nu}{\tilde u^2}\,,\nonumber\\
R_{\mu\nu} & = & \tilde g_{\mu\nu} - Q_{\mu\nu}\,.
\eeqa
In terms of the variables $V_{T,L}$ we have defined in \Eq{VTLk},
\beq
\label{LeffpiTL}
L_{eff} = -(k^2 - \pi_T)V^\ast_T\cdot V_T 
+ (k^2 - \pi_L)V^\ast_L V_L\,.
\eeq

From \Eq{L2withoutVk}, the equation for $\rho$ is precisely the equation
given in the second line of \Eq{VLrhoeqs}. Solving it for $\rho$ and
substituting the solution back in \Eq{L2withoutVk} gives
\beq
\label{L2VTL}
L^{(2)}(k) = -(k^2 - m^2_V)V^\ast_T\cdot V_T +
(k^2 - m^2_V)V^{\ast}_L V_L -
\frac{(2\mu m_V\sqrt{-\tilde u^2})^2}
{k^2 - m^2_1 - \frac{4\mu^2(k\cdot u)^2}{k^2}}V^\ast_L V_L\,.
\eeq
Comparing \Eqs{LeffpiTL}{L2VTL}, we then identify
\beq
\pi_L = m^2_V + 
\frac{(2\mu m_V\sqrt{-\tilde u^2})^2}
{k^2 - m^2_1 - \frac{4\mu^2(k\cdot u)^2}{k^2}}\,,
\eeq
which using \Eqss{omegakappadef}{utilde2}{rhoeffectivemass}
can be written in the form
\beq
\label{piL}
\pi_L = m^2_V +
\frac{4\mu^2 m^2_V \kappa^2}{k^2(k^2 - m^2_{\rho}) - 4\mu^2\kappa^2}\,.
\eeq

It is simple to verify that the dispersion relation equation
\beq
\label{dreqpiL}
k^2 - \pi_L = 0\,,
\eeq
gives again \Eq{dreq1}, as it should be. While this is reassuring,
the virtue of \Eq{piL} is that it gives us an expression
for the longitudinal dielectric function, as function of $\omega$ and $\kappa$,
\beq
\epsilon_\ell = 1 - \frac{\pi_L}{k^2}\,,
\eeq
which allows to study also the response of the system to external fields.

In this context we would like to emphasize the following.
What we have done is to find the modes that diagonalize the
scalar-photon part of the Lagrangian, in the BE background.
The dispersion relations that we have written in \Eq{gbecdr}
are the dispersion relations of those modes.

In reality, as we have mentioned, the BE condensate must be superimposed
on an oppositely charged background component.
It would be a multi-component system, of which one of them is the BE condensate.
Both components contribute to the total $\pi_L$ (as well as the dielectric
constant), and therefore the dispersion relations of the propagating plasma
modes have contributions from both components. Physically
the propagating modes are combinations of the BE modes and the plasma modes
of the background component.

In systems in which the dominant contribution to $\pi_L$ arises from the
BE component [which we have written in \Eq{piL}], then the dispersion
relation of the propagating photon (obtained by solving \Eq{dreqpiL})
is given by the  dispersion relation of the BE modes that we have written
in \Eq{gbecdr}.

In more general cases, $\pi_L$ would have an additional contribution besides
the BE contribution of \Eq{piL}, depending on the composition and conditions
of the background component, and the dispersion relations of the propagating
modes would be correspondingly different from those given in \Eq{gbecdr}.

The main result is that, in such a multicomponent system,
\Eq{piL} represents the contribution to $\pi_L$ (and whence to the dielectric
constant) due to the BE component, which can be used to study the implications
for the optical properties of the system, which are
different from what they would be if the system does not have the BE component.

\section{Mixing}
\label{sec:mixing}

Besides the dispersion relation and the polarization tensor, another
important factor is the $\rho-V_L$ mixing in the eigenmodes.
Once the solutions $\omega_{\pm}$ are given, these are determined
from \Eq{VLrhoeqs}. From \Eq{gbecdr} for the dispersion relations
it follows that for $\kappa = 0$,
\beqa
\omega^2_{+}(0) & = & m^2_{>}\,,\nonumber\\
\omega^2_{-}(0) & = & m^2_{<}\,,
\eeqa
where $m_{>}$ ($m_{<}$) is the greater (lesser) of $m_{\rho}$ and $m_V$.
In what follows we consider the case $m_{\rho} > m_V$. The opposite case
can be treated similarly.
\begin{figure}
\begin{center}
\epsfig{file=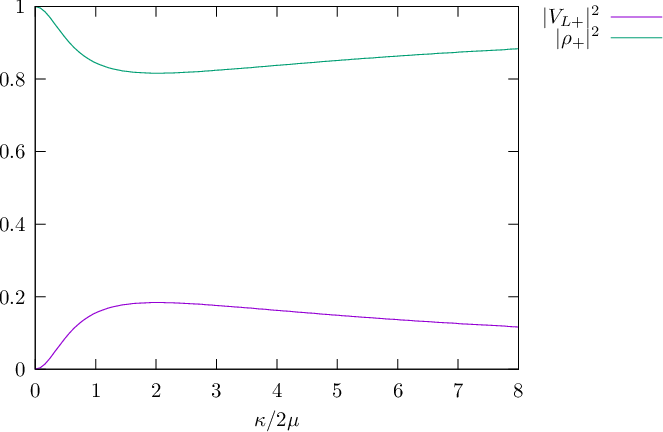,bbllx=166,bblly=291,bburx=484,bbury=499}
\end{center}
\caption[] {
  Plots of the mixing coefficients  $\rho_{+}, V_{L+}$ defined
  in \Eq{mixingcoefficients}, as functions of $\kappa$.
  We have set $M = 2$, and $\Delta = 1$.
  \label{fig:mixingp}
}
\end{figure}
\begin{figure}
\begin{center}
\epsfig{file=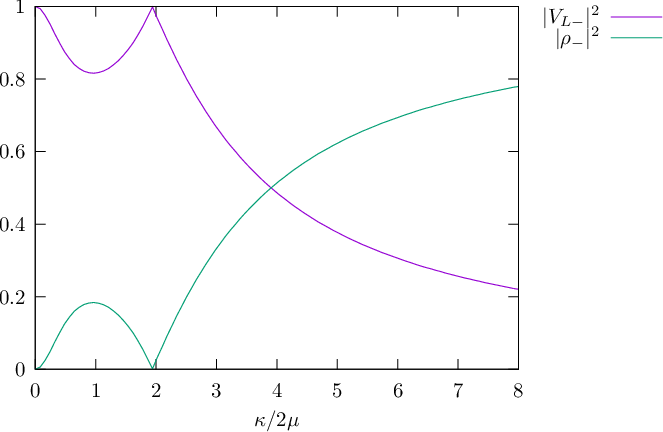,bbllx=166,bblly=291,bburx=484,bbury=499}
\end{center}
\caption[] {
  Plots of the mixing coefficients  $\rho_{-}, V_{L-}$ defined
  in \Eq{mixingcoefficients} as functions of $\kappa$.
  We have set $M = 2$, and $\Delta = 1$.
  \label{fig:mixingm}
}
\end{figure}

A convenient way to write the elements of the mode eigenvectors in this
case is
\beqa
\label{mixingcoefficients}
\left(\begin{array}{c}
  \rho\\ V_L
\end{array}\right)_{-} & = & \frac{1}{\sqrt{N_{-}}}
\left(\begin{array}{c}
  m_V\gamma_{-} \\[12pt] \Lambda + \gamma^2_{-}
\end{array}\right)\,,
\nonumber\\[12pt]
\left(\begin{array}{c}
  \rho\\ V_L
\end{array}\right)_{+} & = & \frac{1}{\sqrt{N_{+}}}
\left(\begin{array}{c}
  \Lambda \\[12pt] -m_V\gamma_{+}
\end{array}\right)\,,
\eeqa
where
\beqa
\label{gammaLambda}
\gamma_{\pm} & = & \frac{2\mu\kappa}{\sqrt{k^2_{\pm}}}\,,\nonumber\\
k^2_{\pm} & = & \omega^2_{\pm} - \kappa^2 = \frac{1}{2}(m^2_{\rho} + m^2_V) \pm
\left[\frac{1}{4}(m^2_{\rho} - m^2_V)^2 + 4\mu^2\kappa^2\right]^{\frac{1}{2}}\,,
\nonumber\\
\Lambda & = & \frac{1}{2}(m^2_{\rho} - m^2_V) +
\left[\frac{1}{4}(m^2_{\rho} - m^2_V)^2 + 4\mu^2\kappa^2\right]^{\frac{1}{2}}\,,
\eeqa
and
\beqa
N_{-} & = & (\Lambda + \gamma^2_{-})^2 + m^2_V |\gamma_{-}|^2\,,\nonumber\\
N_{+} & = & \Lambda^2 + m^2_V\gamma^2_{+}\,.
\eeqa
With this convention, in the limit $\kappa = 0$ these solutions reduce to
\beqa
\label{omegaminusspinor}
\left(\begin{array}{c}
  \rho\\ V_{L}
\end{array}\right)_{-} & = &
\left(\begin{array}{c}
  0\\1
\end{array}\right)\,,\\
\label{omegaplusspinor}
\left(\begin{array}{c}
  \rho\\ V_L
\end{array}\right)_{+} & = &
\left(\begin{array}{c}
  1\\0
\end{array}\right)\,,
\eeqa
when $m_{\rho} > m_V$. Plots of the mixing coefficients
$|\rho_{\pm}|^2$ and $|V_{L\pm}|^2$ is shown in \Fig{fig:mixingp} and
\Fig{fig:mixingm}. For the plots we take $M^2$ and $\Delta$
as indicated in the caption, and we consider the case $m_{\rho} > m_V$,
so that, in particular, $(m_V/2\mu)^2 = \frac{1}{2}(M^2 - \Delta)$.
The plots do not vary significantly for other values of $M^2$ and $\Delta$.

As already indicated above, at $\kappa = 0$ the $\omega_{-}$ solution
corresponds to a pure $V_L$,
\beq
\left(\begin{array}{c}
  \rho\\ V_L
\end{array}\right)_{-} =
\left(\begin{array}{c}
  0\\1
\end{array}\right)\qquad (\kappa = 0)\,.
\eeq
At very large values of $\kappa$, it follows from \Eq{mixingcoefficients}
that the $\rho$ component dominates in both branches, which is
reflected in the plots in \Figs{fig:mixingp}{fig:mixingm}.

On the other hand, the dispersion relation $\omega_{-}$ satisfies
\beq
k^2_{-} = 0\,,
\eeq
for $\kappa = \kappa_0$, where
\beq
\kappa_0 = \frac{m_V m_{\rho}}{2\mu}\,.
\eeq
It is easy to see from \Eq{gammaLambda}, that $\gamma_{-}$
increases relative to $\Lambda$ as $k^2_{-} \rightarrow 0$,
and therefore from \Eq{mixingcoefficients} it follows that
for $\kappa = \kappa_0$ the $(-)$ mode is a pure $V_L$
\beq
\left(\begin{array}{c}
  \rho\\ V_L
\end{array}\right)_{-} =
\left(\begin{array}{c}
  0\\1
\end{array}\right)\qquad (\kappa = \kappa_0)\,.
\eeq
Moreover, $k^2_{-}$ is positive for $\kappa < \kappa_0$, but it is
negative for $\kappa > \kappa_0$. Therefore, for $\kappa > \kappa_0$, we have
\beq
\gamma_{-} = -\frac{2i\mu\kappa}{\sqrt{|k^2_{-}|}}\,,
\eeq
and then the $\omega_{-}$ eigenspinor is of the form
\beq
\left(\begin{array}{c}
  \rho\\ V_L
\end{array}\right)_{-} = \frac{1}{\sqrt{N_{-}}}
\left(\begin{array}{c}
  -i m_V|\gamma_{-}| \\[12pt] \Lambda - |\gamma_{-}|^2\,.
\end{array}\right)\qquad (\kappa > \kappa_0)\,.
\eeq
The situation here resembles the circular polarization of a propagating
photon, or more precisely in this case, elliptic polarization,
\beq
e_{\pm} = 
\left(\begin{array}{c}
  \pm ia\\ b
\end{array}\right)\,.
\eeq
Thus in the present case, if $\kappa > \kappa_0$, the $\omega_{-}$ mode
oscillates between the two \emph{flavors} ($V_L$ and $\rho$), as it
propagates, and at some spacetime points it is a pure $V_L$ and at other
points a pure $\rho$.

\section{Conclusions and Outlook}
\label{sec:conclusions}

We have presented a model for the propagation of photons in a medium
that contains a Bose-Einstein (BE) condensate of a charged scalar field.
The idea is based on the model recently proposed to treat
the propagation of fermions in a BE condensate, extended by coupling the scalar
field with the electromagnetic field in the usual way. Our goal has been
to study the effects of the symmetry breaking on the photon dispersion
relations and the optical properties of the system.
To this end, the propagating modes were identified, and the
corresponding dispersion relations were determined, in \Section{sec:dr}.
As we have shown there, two of the modes correspond
to the transverse photon polarizations, and their dispersion relations
have the usual form of the transverse photons in a plasma. There are
two other modes, to which we refer as the $(\pm)$ modes, that are combinations
of the longitudinal photon and the massive scalar field.
Some salient aspects of the results obtained for the dispersion
relations of the $(\pm)$ modes were discussed in \Section{sec:discussion}.
The dispersion relation of the $(+)$ mode is always greater than the momentum,
and increases steadily as the momentum increases.
The behavior of the dispersion relation for the $(-)$
mode is very different. On one hand, it decreases as the momentum
increases in a given momentum range,
and therefore the corresponding group velocity is negative in that range.
In another range, for values of momentum above a given threshold, it increases
steadily, but such that it remains smaller than the
momentum as it increases, akin to a medium with index of refraction
greater than 1. In \Section{sec:discussion}
we also considered the non-relativistic limit of the dispersion
relations, and discussed some aspects of the results
in particular for the $(-)$ mode. As shown there,
under some conditions the dispersion
relation of the $(-)$ mode reduces to the form
of the dispersion relation that has been obtained in
non-relativistic models of the BE condensation of charged
scalars\cite{2prb,3prb}.
As a complement, in \Section{sec:poltensor} we gave the expression for the
contribution to the photon polarization tensor due to the BE condensate.
This provides an alternative way to obtain the dispersion
relations, but more importantly it has the virtue that it gives the
corresponding contribution to the dielectric function of the system,
which is useful on its own merits, e.g. for studying
also the response of the system to external fields. Another
useful result is the determination of the mixing parameters of the $(\pm)$
modes presented in \Section{sec:mixing}, which can be used to obtain
the corrections to the dispersion relations
(e.g., imaginary parts due the damping effects) and/or the effects of
scattering, due to the interactions with the excitations of the system.

In summary, altogether, the results we have obtained 
show that the BE condensate model we have considered
has some particular non-trivial optical properties.
Our aim has been to analyze them, in the context of the model,
with the expectation that the model and the results obtained can be useful
in some of the specific application contexts considered in the references
cited.

%\begin{acknowledgments}
The work of S. S. is partially supported by 
DGAPA-UNAM (Mexico) PAPIIT project No. IN103522.
%\end{acknowledgments}

\appendix

\section{Justification of \Eq{xkcorrespondence}}
\label{sec:app:xkcorrespondence}

The justification of \Eq{xkcorrespondence} is the following.
For any field we write
\beq
a(x) = \int\frac{d^4k}{(2\pi)^4} a(k) e^{-ik\cdot x}\,.
\eeq
If $a(x)$ is real (Hermitian), then
\beq
a(-k) = a^\ast(k)\,.
\eeq
Now consider something like (an action) involving two real fields
\beq
S = \int d^4x \left[
  \frac{1}{2} a^2(x) + \frac{1}{2} b^2(x) + \alpha b(x)a(x)
  \right]\,.
\eeq
What is the Lagrangian density in momentum space? Using the transformation of
$a$ and $b$,
\beqa
S & = & \int\frac{d^4k}{(2\pi)^4}\int\frac{d^4k^\prime}{(2\pi)^4}
\left[\frac{1}{2} a(k^\prime)a(k) + \frac{1}{2} b(k^\prime)b(k) +
  \alpha b(k^\prime)a(k)\right]\delta(k + k^\prime)\nonumber\\
& = & \int\frac{d^4k}{(2\pi)^4} \left[
  \frac{1}{2} a(-k)a(k) + \frac{1}{2} b(-k)b(k) + \alpha b(-k)a(k)
\right]\,.
\eeqa
But we would not obtain the correct equations if we just use
\beq
L(k) = \frac{1}{2} a(-k)a(k) + \frac{1}{2} b(-k)b(k) + \alpha b(-k)a(k)\,,
\eeq
which is equal to
\beq
L(k) = \frac{1}{2} a^\ast(k)a(k) + \frac{1}{2} b^\ast(k)b(k) +
\beta b^\ast(k)a(k)\,.
\eeq
In particular, notice that this $L(k)$ is not real (Hermitian). The correct
procedure is to notice that, since the first term mixes $a(k)$ with $a(-k)$,
and similarly with $b(k)$, we must include both also in the $\alpha$ term.
Therefore we must write
\beq
S = \int_{k > 0}\frac{d^4k}{(2\pi)^4}\left[
  \frac{1}{2} a(-k)a(k) + \frac{1}{2} b(-k)b(k) + \alpha b(-k)a(k) +
  (k \leftrightarrow -k)
  \right]\,,
\eeq
and therefore the correct Lagrangian density to use in momentum space is
\beq
L(k) = a^\ast(k)a(k) + b^\ast(k)b(k) + \alpha\left(b^\ast(k)a(k) + c.c\right)\,,
\eeq
with the understanding that the complete action is
\beq
S = \int_{k > 0}\frac{d^4k}{(2\pi)^4} L(k)\,.
\eeq
%

%\bibliographystyle{ieeetr}
%\bibliography{main}

\begin{thebibliography}{99}
\bibitem{footnote1}
  The presence of $\mu$ in the Lagrangian can be eliminated by letting
  % 
  \beq
  A_\mu = A^\prime_\mu - \frac{1}{q}v_\mu\,.
  \eeq
  %
  However, doing this transformation of $A_\mu$ would restore
  the $\mu$ in the partition function, which defeats the purpose
  of using the transformed field $\phi^\prime$.
  Of course this is nothing more than the statement of
  invariance under a gauge transformation
  %
  \beqa
  \phi & = & e^{iq\Lambda}\phi^\prime\,,\nonumber\\
  A_\mu & = & A^\prime_\mu - \partial_\mu\Lambda\,.
  \eeqa
\end{thebibliography}

\end{document}